# Quantum transport in oxide nanostructures

Cheng Cen[1], Daniela F. Bogorin[1], Vanita Srinivasa[1], Jeremy Levy[1*]

[1]*Department of Physics and Astronomy, University of Pittsburgh, 3941 O'Hara St., Pittsburgh, PA 15260, USA.*

## Abstract

We describe magnetotransport experiments performed on Hall crosses made from quantum wires at $LaAlO_3/SrTiO_3$ interfaces. Shubnikov-de Haas oscillations measured in a 14-nm wide structure exhibit modulations that are consistent with spin-orbit coupling or valley degeneracies. Hall measurements performed on the 6 nm-wide Hall cross reveal dissipative coupling to magnetic phases. Hall plateaus are observed that deviate significantly from two-dimensional quantum Hall counterparts. Non-monotonic dips in the Hall resistance are attributed to one-dimensional confinement and spin-orbit coupling.

[*] E-mail: jlevy@pitt.edu

The conducting interface between insulating perovskite oxides LaAlO$_3$ and SrTiO$_3$ [1] has sparked wide interest due to the multifaceted nature of SrTiO$_3$. High interface mobilities $\mu$~$10^4$ cm$^2$/Vs have been reported [2, 3], along with superconductivity [4], Shubnikov-de Haas (SdH) oscillations [2, 3], anomalous magnetic behavior [5, 6], large spin-orbit coupling [7, 8], strong correlations [9] and phase-coherent transport [10].

While materials advances have led to reports of integer and fractional quantum Hall (IQH and FQH) effects in ZnO-Mg$_x$Zn$_{1-x}$O oxide heterostructures [11, 12], the observation of comparable effects in the LaAlO$_3$/SrTiO$_3$ system has remained elusive. One challenge pertains to the relatively large two-dimensional (2D) carrier density ($n$~$10^{13}$-$10^{14}$ cm$^{-2}$), roughly two orders of magnitude higher than for traditional quantum Hall systems. Others concern the existence of degenerate (and possibly universal [13]) bands at the interface, and questions about the precise nature of localization at the oxide interface. Prior measurements in high-mobility LaAlO$_3$/SrTiO$_3$ heterostructures and modulation-doped SrTiO$_3$ indicate an order of magnitude difference between the carrier density quantified by Hall effect measurements and the corresponding frequency of SdH oscillations. One proposed explanation [14, 15] is that electrons phase-separate into subbands with varying mobility, and only the highest mobility electrons contribute to the observed oscillations (which appear with relatively small amplitude). The origin of carriers within the two-dimensional electron gas (2DEG) has also been controversial; some reports suggest that intermixing is predominantly responsible for electronic doping of the interface [16], rather than "polar catastrophe" [1].

Quasi-one-dimensional (q-1D) confinement of electrons using conductive atomic-force microscopy (c-AFM) [17] offers new ways for controlling electronic properties at the LaAlO$_3$/SrTiO$_3$ interface. Positive voltages applied to the c-AFM tip are believed to locally charge the top LaAlO$_3$ surface [18, 19], creating conducting states via surface modulation doping. The process is reversible and has been used to create metallic nanowires as narrow as 2 nm [18], which is below the Fermi wavelength and mean-free path at low temperatures. Lateral spatial confinement provided by the nanowire writing process produces nanowires with significantly lower effective carrier densities ($n<3\times10^{12}$ cm$^{-2}$), qualitatively changing the behavior compared with two-dimensional counterparts. The conducting nanostructures are created from an insulating interface, thereby ruling out the possibility of electron doping via interdiffusion.

The LaAlO$_3$/SrTiO$_3$ heterostructures are grown by pulsed laser deposition, at 770 °C in an O$_2$ pressure of $6\times10^{-5}$ mbar [20]. Ohmic contacts to the interface are created using methods described in Ref. [20]. Measurements are performed on two nanowire structures, labeled A and B, created on different heterostructures using nominally identical growth conditions.

Structure A is a Hall bar (Fig. 1(a)) constructed from nanowires with measured width $w_A$=14 nm, determined by cutting a nanowire that written in the same area under similar conditions [17]. Magnetotransport measurements are performed in a $^3$He refrigerator. After cooldown, three of the Hall bar leads ($L_2$, $L_3$, $L_6$) became highly insulating. During cooldown, the resistance of the main channel ($L_1$-$L_4$) increased by two orders of magnitude but remained conductive with a resistance $R_{14}$~100 MΩ at $T$=250 mK (Fig. S1 [21]). These increases in resistance indicate the existence of potential barriers along the channels, which are overcome at room temperature via thermal activation [22].

The L-shaped $L_4$- $L_5$ section had the lowest impedance at 250 mK ($R$=3.08 MΩ), where two-terminal magnetoresistance measurements are performed. The magnetoresistance is asymmetric with respect to magnetic field direction (Fig. 1(b)), similar to those reported elsewhere [23]. After subtracting a quadratic background $R_C$, pronounced SdH oscillations are revealed for $\Delta R = R - R_C$ (Fig. 2).

In Fermi liquids, the temperature dependence of SdH oscillations can be modeled by the Lifshitz-Kosevich (L-K) relation:

$$\Delta R = 4R_0 \sin(2\pi\left(\frac{B_1}{B} - \frac{1}{2}\right) + \frac{1}{4}) \exp(-2\pi^2 k_B T_D / \hbar\omega_C) \frac{2\pi^2 k_B T / \hbar\omega_C}{\sinh\left(2\pi^2 k_B T / \hbar\omega_C\right)} \quad (1)$$

where $T_D$ is the Dingle temperature, $R_0$ is the amplitude of the resistance oscillations, $k_B$ is the Boltzmann constant, $\omega_c = eB/m^*$ is the cyclotron frequency, $B_1 = nh/eN_v N_s$ is the oscillation frequency, $N_v$ is the valley degeneracy, $N_s$ is the spin degeneracy, and $m^*$ is the effective mass. A nonlinear least-squares fit yields values $m^*/m_0 = 1.1$ ($m_0$ is the bare electron mass), $T_D = 1.1$ K, $R_0 = 5.2$ kΩ and $B_1 = 36.5$ T (Fig. 2). Assuming a Landé $g$ factor of ~2, Zeeman splitting in this system is very close to the Landau splitting and a spin degeneracy of 2 is expected, with which a sheet carrier density of $n = 1.7 \times 10^{12}$ cm$^{-2}$ is calculated. While the overall fit to the L-K form is good for intermediate magnetic field values, there are clear deviations. A large (~π) phase shift is observed in the SdH oscillations, which takes place at around 6T (Fig. 2). In the magnetic field range between 13T and 18T, additional structure is observed.

Eq. (1) was originally derived for systems in which single band is occupied. The observed modulation could arise from valley degeneracies or from interface-derived effects such as Rashba spin-orbit coupling. In either case, the ratio between the subband splitting Δ and Landau level splitting $\Delta_L$ can be written as $\Delta/\Delta_L = N + x$, where $N$ is an integer and $0 < x <= 1$. As the magnetic field increases, a condition $x > 1/2$ can be achieved, in which case the maxima in the density of states occur between two Landau levels, which can produce a phase shift up to π. At higher field, when $x\Delta_L$ becomes larger than the Landau level broadening, Landau levels from different subbands can be well resolved and additional SdH oscillation periods will appear. No quantum oscillations were observed for the channel connecting $L_1$-$L_4$ or $L_1$-$L_5$, presumably because of the too large background resistance.

It is remarkable that periodic oscillations are observed at all. In the field range of 3 T-18 T, the cyclotron diameter $l_c$ varies between 70 nm and 12 nm, which is much larger or comparable to the conducting channel width (14 nm). In this regime, magnetoelectric subbands due to the lateral electrostatic confinement, instead of purely magnetic Landau levels, are expected to form [24, 25]. In that case, the separation between two magnetoelectric subbands is no longer a linear function of $B$, which contradicts our observation. Since the Fermi wavelength is larger than the channel width, geometric effects in the ballistic transport regime can be ruled out. One possibility is that electrons in the junction area, where $L_1$-$L_4$, $L_3$-$L_5$ overlap, contribute dominantly to the SdH oscillations. In this scenario, transmission probabilities are strongly affected by the electron occupancy of bound or scattering states that are localized at the junction.

Structure B is a Hall bar constructed from even narrower nanowires of width $w_B$=6 nm (Fig. 3(a)). Structure B is cooled to the base temperature of a dilution refrigerator $T$=20 mK. Between $L_1$ and $L_4$, the resistance decreases from 1.5MΩ to 4.7 kΩ upon cooling. The resistances among $L_2$, $L_3$, $L_5$ increased to ~200 MΩ, again indicating the presence of large but surmountable potential barriers along the channel, whereas $L_6$ became highly insulating. The Hall resistance was measured by sourcing a current ($I$=4.8 nA at frequency $f$=9 Hz) between leads $L_3$ and $L_5$ and measuring the Hall voltage $V_H$ between leads $L_1$ and $L_4$. The (complex) Hall resistance $R_H$=$V_H$/$I$ (Fig. 3(b)), shows strong departures from linear behavior, and distinct asymmetries between positive and negative magnetic field directions. In addition, significant phase shifts between the applied current and Hall voltage are observed in the range -3T < $B$ <3T, with corresponding anomalies in the Hall response. Decomposition of the Hall resistance signal into symmetric and antisymmetric components $2R_H^{\pm}(H) \equiv R_H(H) \pm R_H(-H)$ reveals that the phase shifts are associated exclusively with the symmetric (anomalous) Hall component $R_H^+$ (Fig. 3(c, d)). Several features are observed over a cascading range of magnetic field scales (Fig. 3(b), middle inset). The time scale for the observed phase response is comparable to that of magnetic states previously reported for this system [5]. A strong asymmetry in the slope (almost a factor of two) between positive and negative field values will inevitably lead to error in the estimation of the carrier density; nonetheless, an average value of $n = 1.4 \times 10^{12} \text{cm}^{-2}$ can be obtained from the slope of $R_H^-$ at low field. At larger magnetic fields, distinct Hall plateaus are observed, both for positive and negative fields. The separation between plateaus generally follows $\Delta \frac{1}{\mu_0 H} = \frac{eN_s N_v}{n^* h}$, where $n^* = 3.0 \times 10^{12}$ cm$^{-2}$. The overall agreement between $n$ and $n^*$ suggests that a single subband is occupied. Hall plateaus cease to exist in lower magnetic field, which is probably due to the approaching of the quasi-one-dimensional limit as the cyclotron radius gets larger [26].

The observed Hall plateau resistances deviate significantly from the quantized values of $R_H = \frac{h}{\nu e^2}$. For example, the plateau at around 15T is expected to have a Landau level filling factor of $\nu=4$, however the measured Hall resistance is 6.8 k$\Omega$, which is larger than $\frac{h}{4e^2} = 6.5$ k$\Omega$. Such deviations become more significant at lower field and may be due to the local geometry of the nanometer-sized Hall cross and the discrete nature of electronic states localized at the junction [27, 28]. From the perspective of the Landauer formula $G = (e^2/h) \sum_n T_n$ [29, 30], these states modify the transmission probabilities $T_n$ (where $n$ labels the conducting mode or quantum channel) that give rise to the usual quantized conductance in units of $e^2/h$ expected for a wider straight wire segment and lead to corresponding deviations in the measured Hall resistance. The existence of sharp Hall plateaus indicates a long phase coherence time that emerges, despite the low dimensionality, when a single SrTiO$_3$ subband is occupied. The magnetoresistance (Fig. S2 [21]) measured between $L_1$ and $L_4$ exhibits an asymmetric profile similar to Figure 1(b). However, no SdH oscillations are observed.

One remarkable feature observed at large (+15 T and -17 T) magnetic field strengths is the appearance of non-monotonic dips in the Hall resistance (Figure 3(b), left and right insets). A decrease in Hall resistance is unusual because it represents an increase in the number of edge states with increasing magnetic field. Such effects are unique in quasi-one-dimensional systems and have been predicted theoretically [31] and observed in semiconductor nanowires [32] where strong Rashba spin-orbit coupling was known to be present [33]. Rashba spin-orbit coupling strengths as large as $\alpha = 5 \times 10^{-12}$ eV·m have recently been reported for this system [7, 8], and may be responsible for these features.

In order to explore this possibility, we calculate the quasi-one-dimensional electronic energy subbands for a quantum wire with Rashba spin-orbit coupling as a function of the strength of an externally-applied perpendicular magnetic field $\vec{B} = B\hat{z}$ [25, 34, 35]. The total Hamiltonian describing the system is given by $H = H_0 + H_{so}$, where

$$H_0 = \frac{1}{2m^*}\left[ p_x^2 + \left(p_y + eBx\right)^2 \right] + \frac{1}{2}m^*\omega_0^2 x^2 + \frac{1}{2}g\mu_B B\sigma_z,$$

$$H_{so} = \frac{\alpha}{\hbar}\left[\left(p_y + eBx\right)\sigma_x - p_x\sigma_y\right],$$

(2)

$p_x$ and $p_y$ are the operators for the electron momentum with respect to the lateral confinement (*x*) and propagation (*y*) axes, $m^*$ and $-e$ denote the effective mass and charge of the electron, $g$ is the electron g-factor, $\mu_B$ is the Bohr magneton, and $\left(\sigma_x, \sigma_y, \sigma_z\right)$ are the Pauli spin matrices. The lateral confinement is harmonic with frequency $\omega_0$ and characteristic length scale $l_0 \equiv \sqrt{\hbar/m^*\omega_0}$. Using parameters relevant to the *w*=6 nm Hall cross of Figure 3 ($l_0 = 6$ nm, $m^* = 1.1\, m_e$, $g = 2$) and, assuming that the electron wavefunction is localized primarily within the SrTiO$_3$ layer, we numerically solve the time-independent Schrödinger equation for the energy eigenvalues $E(B)$ as a function of the momentum $k_y = p_y/\hbar$ along the wire. Figs. 4(a) and 4(b) show the lowest few subbands for $\alpha = 5.3\times 10^{-12}$ eV·m and $\alpha = 1.1\times 10^{-11}$ eV·m, respectively, at $B = 15$ T. A distinct local maximum–a spin-orbit gap–appears in the lowest subband for the larger spin-orbit coupling strength (Fig. 4(b)).

The zero-temperature ballistic conductance $G$ of the quantum wire as a function of the Fermi energy $E_F$ can be calculated from the corresponding subband spectra via the relation [31, 36]

$$G(B, E_F) = \frac{e^2}{h}\left\{\sum_{l,i}\theta\left[E_F - E_{\min,i}^{(l)}(B)\right] - \sum_{l,j}\theta\left[E_F - E_{\max,j}^{(l)}(B)\right]\right\}.$$

(3)

Here, $E_{\min,i}^{(l)}(B)$ and $E_{\max,j}^{(l)}(B)$ denote the energies of the $i^{th}$ minimum and the $j^{th}$ maximum, respectively, in the $l^{th}$ energy subband for the magnetic field strength $B$, and $\theta$ is the Heaviside unit step function. According to Eq. (3), subband local maxima can produce dips in the Hall resistance (i.e., increases in $G$) with increasing magnetic field [24, 25]. In Figures 4(c, d), we plot $\mathrm{Sign}(dG/dB)$ as a function of $B$ and $E_F$ for $\alpha = 5.3 \times 10^{-12}\,\mathrm{eV \cdot m}$ and $1.1 \times 10^{-11}\,\mathrm{eV \cdot m}$. The points for which $\mathrm{Sign}(dG/dB) = 1$ correspond to dips in the quantized resistance. For the lower value of $\alpha$ shown (Fig. 4(c)), the nontrivial values of $\mathrm{Sign}(dG/dB) = 1$ appear at magnetic field values well below $B = 15\,\mathrm{T}$. Increasing $\alpha$ tends to shift the distribution of these points to higher magnetic field strengths, reflecting the fact that larger spin-orbit interaction energies are able to compete with larger magnetic field energy splittings. When $\alpha$ is increased to $1.1 \times 10^{-11}\,\mathrm{eV \cdot m}$ (Fig. 4(d)), points with $\mathrm{Sign}(dG/dB) = 1$ appear slightly above the lowest energy minimum at $B = 15\,\mathrm{T}$. Therefore, if the dip in the Hall resistance observed in Figure 3 (b) at 15 T is attributable to Rashba spin-orbit coupling, these calculations imply that $\alpha > 8 \times 10^{-12}\,\mathrm{eV \cdot m}$.

In summary, we report two examples of quantum transport in nanostructures formed at the LaAlO$_3$/SrTiO$_3$ interface. The control over both the effective carrier density and the nanometer structure sizes allows for the tuning of the relative ratio between different energy scales, such as Landau splitting, spin-orbit coupling splitting, lateral confinement potential and magnetic energy. The interplay between these competing interactions portends rich physics in this highly configurable electronic system.

**Acknowledgement**: We thank S. Thiel and J. Mannhart at the University of Augsburg LaAlO$_3$/SrTiO$_3$ for providing samples used for these measurements. Work at NHMFL is performed under the auspices of the NSF, DOE and State of Florida. We thank Tim Murphy at NHMFL for experimental assistance. This work was supported by NSF DMR-0704022, NSF-0948671 and ARO MURI W911NF-08-1-0317. V.S. acknowledges helpful discussions with Y. Pershin, and support from NDSEG and an Andrew Mellon Fellowship.

**Fig. 1.** (a) Schematic for structure A. Two-terminal magnetotransport measurements are performed on the highlighted section between leads $L_4$ and $L_5$. (b) Magnetoresistance of the nanowire at $T=0.25$ K.

**Fig. 2.** Reprocal plots of $\Delta R$ at four temperatures $T=0.25$ K, $T=0.5$ K, $T=1.0$ K, $T=2.0$ K. Also shown is a fit to the Lifshitz-Kosevich relation Eq. (1).

**Fig. 3.** (a) Schematic of structure B. (b) Magnitude and phase of Hall resistance measured at $f=9$ Hz. Insets show several features at low and high magnetic fields. (c) Magnitude and phase of symmetric ($R_H^+$) Hall component. (d) Magnitude and phase of antisymmetric ($R_H^-$) Hall component.

**Figure 4.** (a) Lowest few electronic energy subbands for $\alpha = 5.3\times10^{-12}$ eV·m and $B=15$ T. (b) Lowest few subbands for $\alpha = 1.1\times10^{-11}$ eV·m and $B=15$ T, showing a local maximum in the lowest subband. (c) Plots of points for which $\text{Sign}(dG/dB)=1$ and $\text{Sign}(dG/dB)=-1$ (red and blue points, respectively) as a function of $B$ and $E_F$, and lowest energy minimum of all subbands for each value of $B$ (green squares), for the Rashba spin-orbit coupling strength in (a).

(d) Same as (c), but for the Rashba spin-orbit coupling strength in (b). The arrow indicates points at B~15 T which correspond to dips in the quantized resistance.

(a)

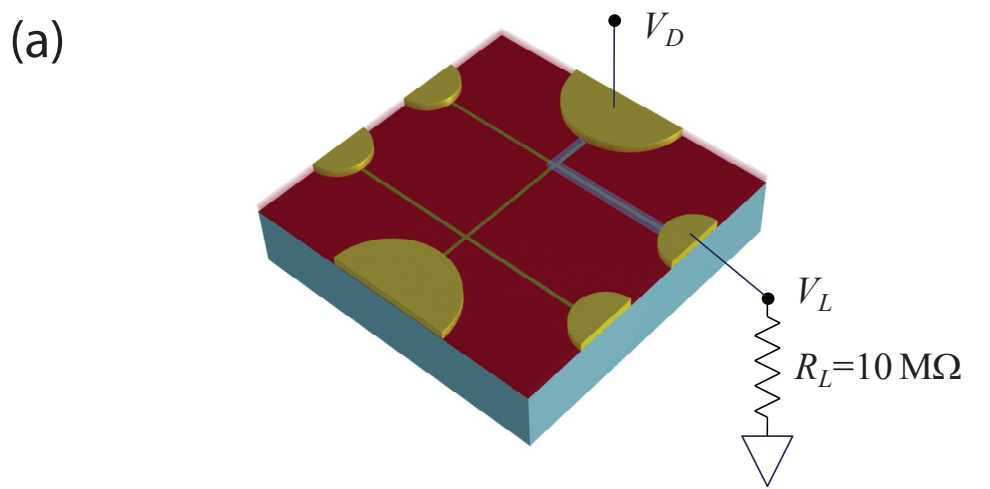

(b)

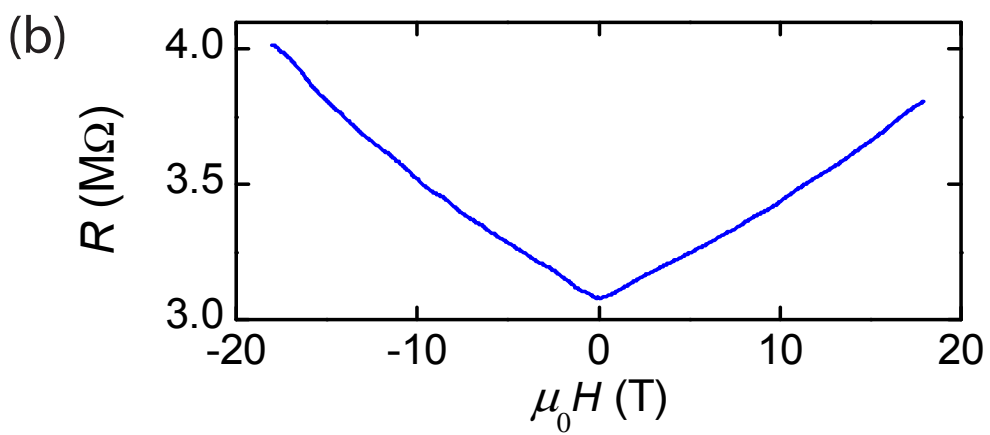

Cen et al, Figure 1

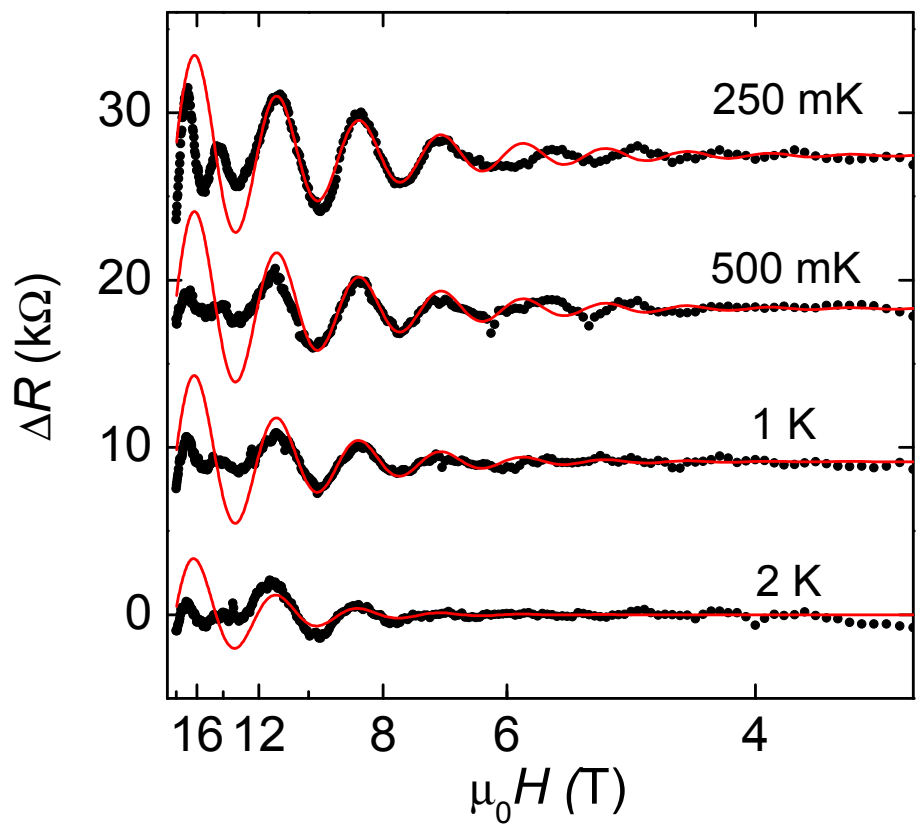

Cen et al, Figure 2

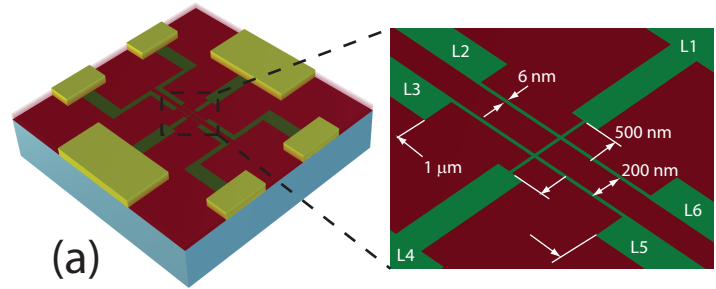
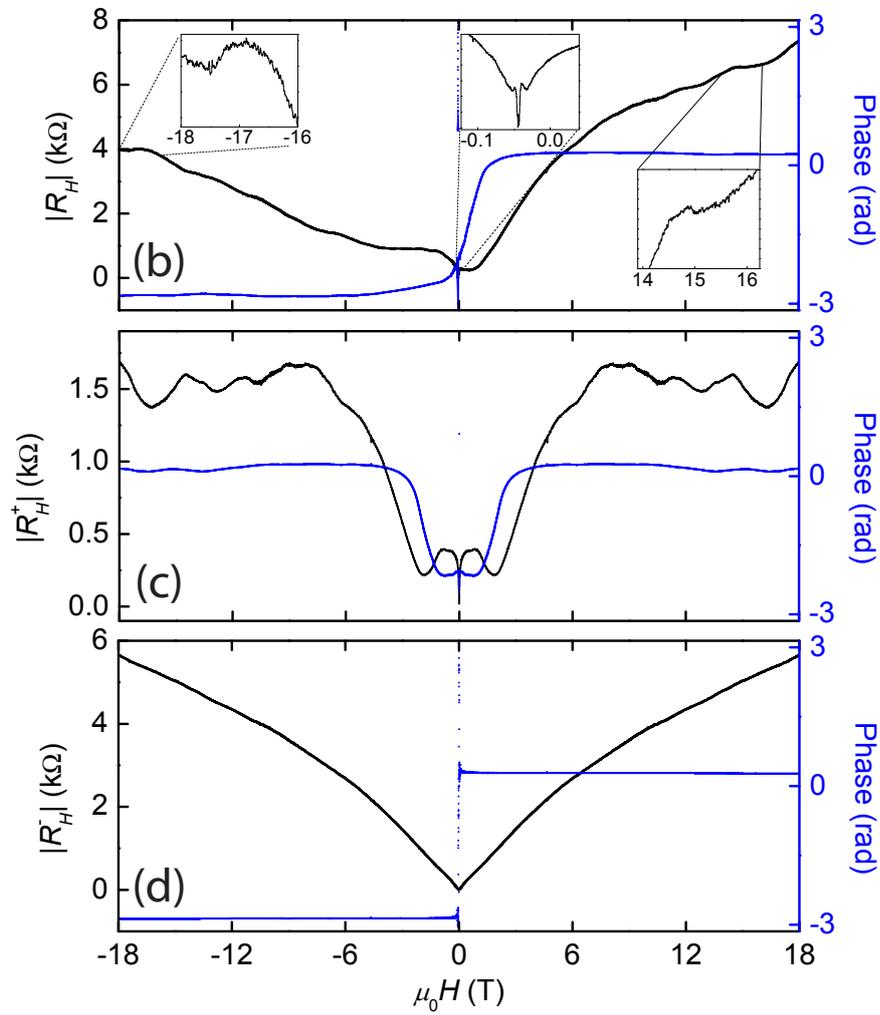

Cen et al Figure 3

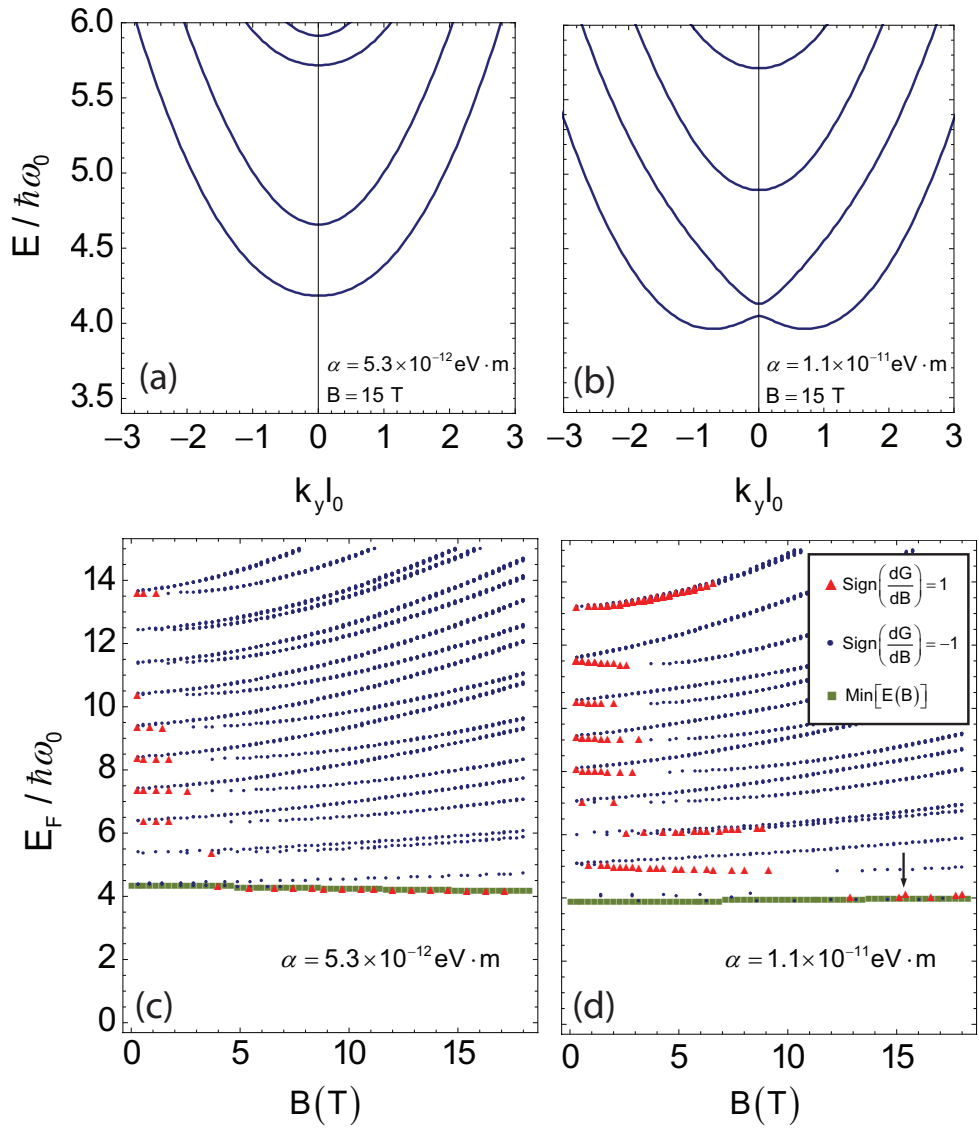

Cen et al, Figure 4

# Quantum transport in oxide nanostructures:

# Supplementary Materials

Cheng Cen[1], Daniela F. Bogorin[1], Vanita Srinivasa[1], Jeremy Levy[1*]

[1]*Department of Physics and Astronomy, University of Pittsburgh, 3941 O'Hara St., Pittsburgh, PA 15260, USA.*

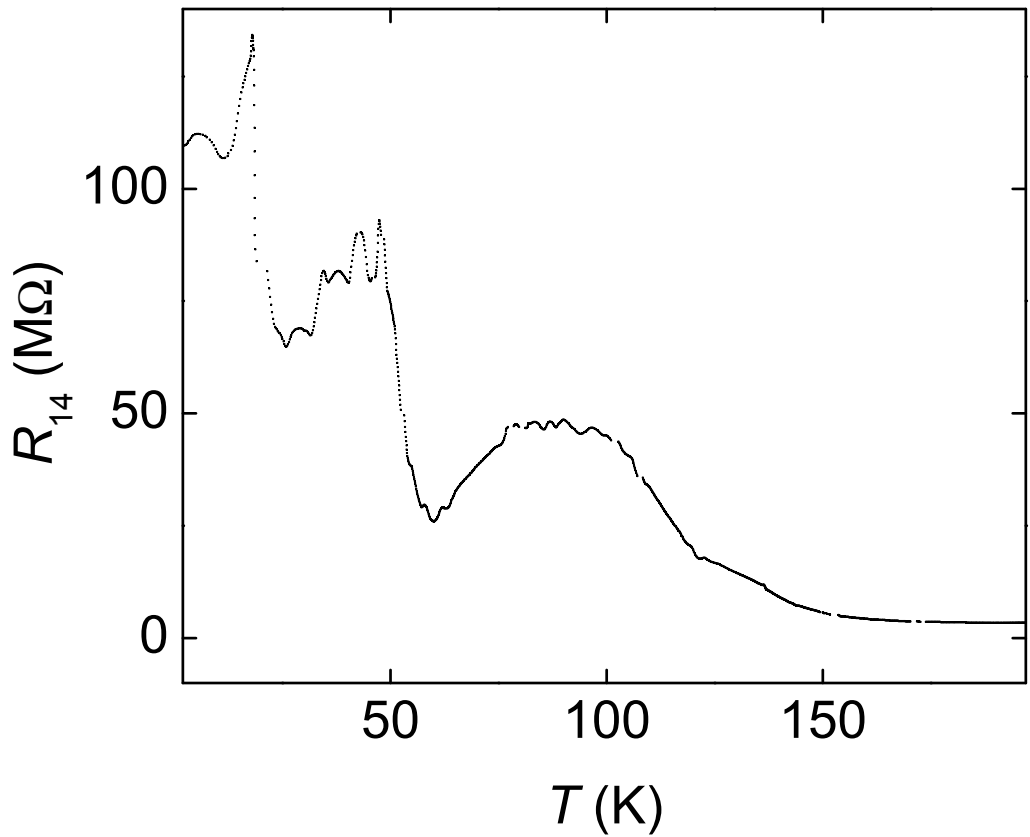

**Figure S1** Two-probe resistance between leads $L_1$ and $L_4$ in structure A measured during the cooldown.

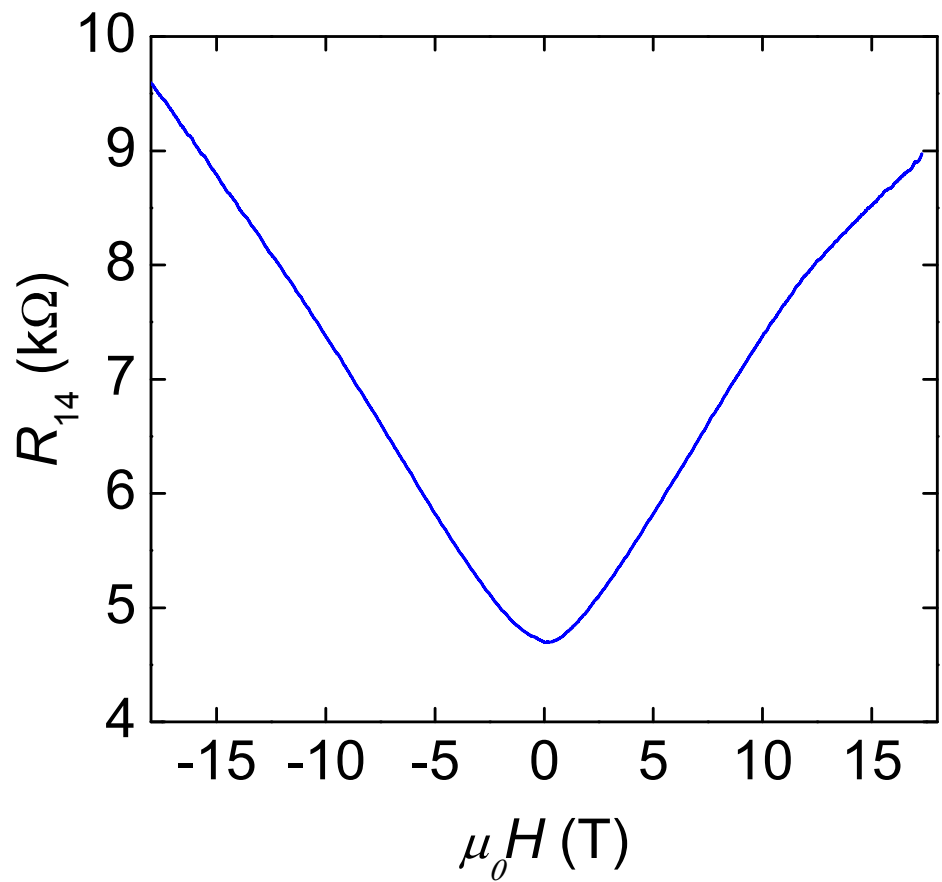

**Figure S2** Two-probe resistance between leads $L_1$ and $L_4$ in structure B measured as a function of magnetic field.